\documentclass{article}
\usepackage{spconf,amsmath,graphicx}
\usepackage{amsfonts}
\usepackage[caption=false,font=footnotesize]{subfig}
\usepackage[T1]{fontenc}
\usepackage{mathtools}    
\usepackage{algorithm}
\usepackage{algpseudocode}
\usepackage{color}

\title{Deep Back Projection for Sparse-View CT Reconstruction}
\name{Dong Hye Ye$^*$, 
Gregery T. Buzzard$^\dag$,
Max Ruby$^\dag$, and 
Charles A. Bouman$^*$ \thanks{C.A. Bouman and G.T. Buzzard were partially supported by NSF CCF-1763896.  
}
}
\address{$^*$School of Electrical and Computer Engineering, Purdue University, West Lafayette, IN, 47907 \\
$^\dag$Department of Mathematics, Purdue University, West Lafayette, IN, 47907}
\begin{document}
%
\maketitle
\begin{abstract}
Filtered back projection (FBP) is a classical method for image reconstruction from sinogram CT data. FBP is computationally efficient but produces lower quality reconstructions than more sophisticated iterative methods, particularly when the number of views is lower than the number required by the Nyquist rate. In this paper, we use a deep convolutional neural network (CNN) to produce high-quality reconstructions directly from sinogram data. A primary novelty of our approach is that we first back project each view separately to form a stack of back projections and then feed this stack as input into the convolutional neural network. These single-view back projections convert the encoding of sinogram data into the appropriate spatial location, which can then be leveraged by the spatial invariance of the CNN to learn the reconstruction effectively. We demonstrate the benefit of our CNN based back projection on simulated sparse-view CT data over classical FBP.  
\end{abstract}
\section{Introduction}
\label{sec:intro}

Computed Tomography (CT) is very important in applications ranging from health care and manufacturing to scientific exploration \cite{sidky2008}. Traditional CT reconstruction algorithms require approximately $n$ views, each with $n$ channels, in order to reconstruct an $n \times n$ image since this ensures Nyquist sampling. However, collecting $n$ views is not always possible or practical. So for example, some security CT scanners use a fixed set of source positions, which leads to sparse views. In manufacturing applications, sparse view collection can reduce acquisition time, which reduces cost.  In scientific applications, it is often not possible to collect a full set of views, particularly when imaging dynamically changing objects.

Regularized iterative methods, such as Model Based Iterative Reconstruction (MBIR), can form high-quality images from sparse views by incorporating into the reconstruction problem a forward model of the physics of the CT scanner together with a prior model of the object being imaged \cite{Thibault07,Zhang14tmi,Jin15}.  However, while progress has been made in speeding up MBIR, it remains computationally expensive due to the nature of iterative optimization, which limits its use in practical applications.

Recently, there has been a great deal of interest in using deep convolutional neural networks (CNN) for image processing tasks. We make no attempt to survey this literature, but a few of many examples include effective uses of CNNs for denoising, image tracking, and object recognition \cite{RonnebergerEtAl,ye2018ei,ZhangEtAl}. More relevant for this paper, applications of CNNs for CT reconstruction have proliferated in recent years, with most methods focused on applying a CNN to a reconstructed image to reduce artifacts/increase quality. Some of these approaches use a CNN as a single-step denoiser, while others train a CNN to act as part of an iterative reconstruction method \cite{Wang,KangEtAl,ye2018icassp,BatenburgKosters,SidkyEtAl,JinEtAl,McCannJinUnser,gupta2018,HanYe,KellyEtAl,Chen2017,SchlemperEtAl}.

The power of CNNs is that they reduce the number of parameters of deep neural networks by imposing spatial invariance.  The challenge of this in applying to sinograms is that the sinogram has information encoded in a spatially nonlocal way. In one approach to overcome this challenge, \cite{AdlerOktem} use CNNs to learn approximate proximal maps for a version of the primal-dual algorithm. However, their approach uses a type of memory in both the primal and dual spaces and multiple learned proximal maps for different steps in the iteration, thus increasing the parameter space significantly. A related approach for MRI is given in \cite{YangEtAl}.

\begin{figure*}[!t]
\centering
\subfloat{\includegraphics[width=7in]{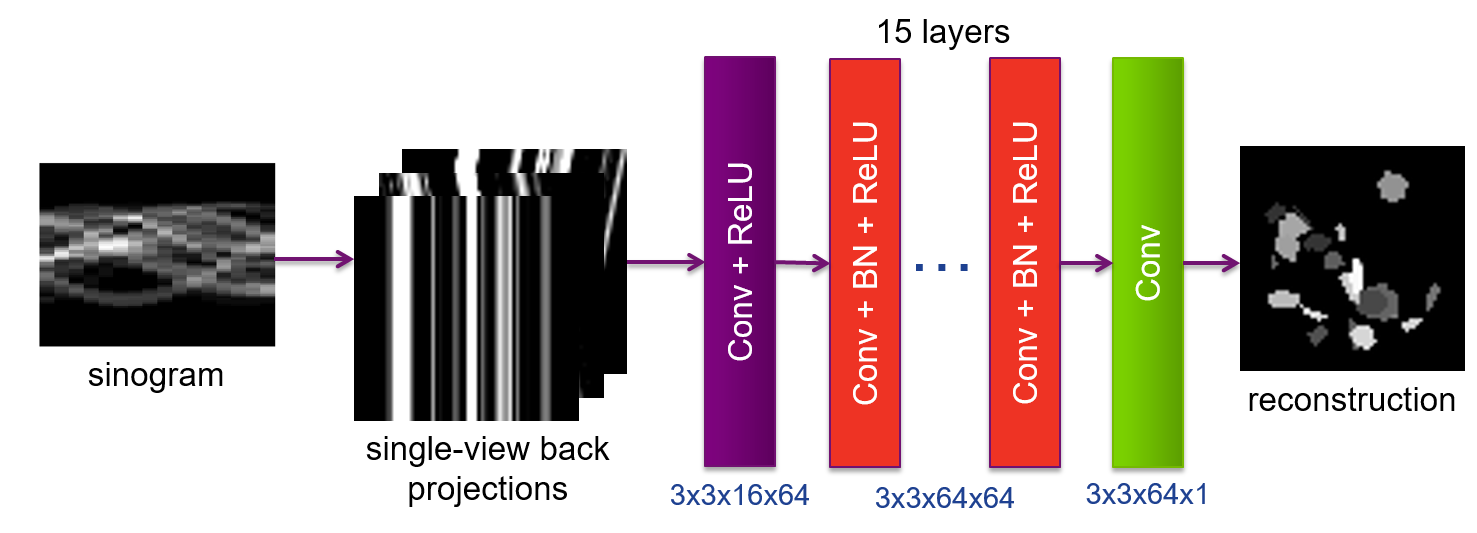}
}
\caption{Overview of Deep Back Projection (DBP): Given a sinogram, we generate single-view projections for all view angles. Multi-layer convolutional neural networks are trained to predict a reconstructed image. Note that we use batch normalizaton (BN) and rectified linear units (ReLU) for efficient training of deep neural networks.}
\label{fig:network}
\end{figure*}

In this paper, we introduce a method called Deep Back Projection (DBP), in which we use a deep convolutional neural network to produce high-quality reconstructions directly from sinogram data. The challenge in this problem is finding a way to reorganize the sinogram data to make it amenable to processing by a CNN. In a primary novelty of our approach, we first create what we call a stacked back projection by back projecting each view separately to form an image of parallel lines of varying intensity (see Fig.~\ref{fig:network}). We then feed this stack as input into the convolutional neural network. This minimal, linear preprocessing step converts the nonlocal information encoded in the sinogram into spatially invariant information in reconstruction space, which can then be leveraged by the spatial invariance of the CNN to learn the full reconstruction effectively with relatively few parameters. 

\section{Deep Back Projection}
\label{sec:method}
In Fig.~\ref{fig:network}, we show the overall framework of our Deep Back Projection (DBP) method. Given a sparsely measured CT sinogram (e.g., 16 view angles), we back project each view separately to form a stacked back projection. Then, we train deep convolutional neural networks to predict reconstructed image from this stack of single-view back projections. For the network architecture, we use a multi-layer convolutional filter with batch normalization and rectified linear units. By feeding single-view projections of unseen testing CT data into this trained network, we can reconstruct the image directly from the sinogram. In the following, we describe each component of our DBP in more detail. 

\subsection{The Stacked Back-Projection Tensor}
Let $y$ denote a set of 2D CT data with $n$ channels and $m$ views.
Then $y$ can be viewed as an $n \times m$ sinogram image, where the $i^{th}$ column corresponds to an individual view projection $y_i$ taken at angle $\theta_j$.
So then $y=[y_1, \cdots , y_m]$ where each element of the vector $y_j\in \Re^n$ is essentially the integral of the pixel intensities over a line with angle $\theta_j$ and an offset from center that depends on the channel. 

\begin{figure*}[!t]
\centering
\subfloat{\includegraphics[width=7in]{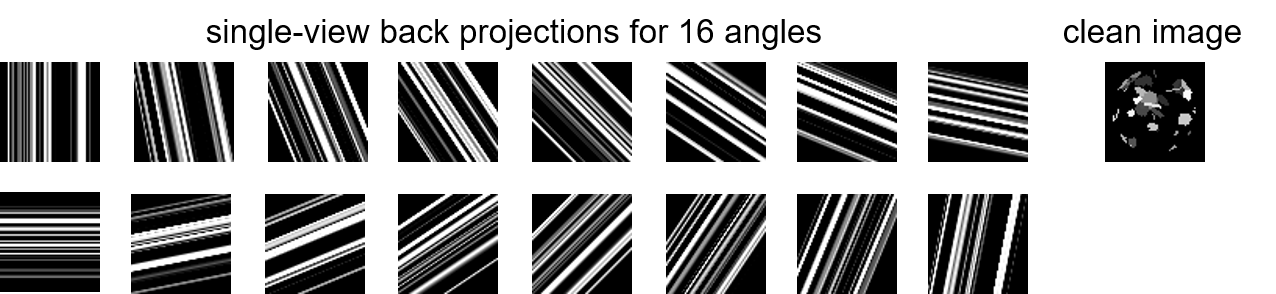}
}
\caption{Example of dataset. Radon transform is applied to the clean synthetic image with 16 view angles to generate the CT sinogram. Single-view back projections from the sinogram are stacked to form the input for a convolutional neural network. Note that each single-view back projection shows lines with constant intensity with respect to the corresponding view angle.}
\label{fig:tensor}
\end{figure*}

Each view projection, $y_i$ is then back projected along the corresponding angle to form an $n \times n$ image of parallel lines. 
Since this back projection operator is linear, the single-view back projected image $Z_j\in\Re^{n\times n}$ can be computed as
\begin{equation}
Z_j=B_j y_j,
\label{eq:singleviewBP}
\end{equation}
where $B_j$ is a single-view back projection operator at the particular view angle $\theta_j$.
The images $Z_j\in \Re^{n\times n}$ are then stacked to form a single tensor of dimension $n\times n \times m$.
$$
\mathbf{Z} = [ Z_1; \cdots ; Z_m] 
$$
We call $\mathbf{Z}$ the back-projection tensor (see the example in Fig.~\ref{fig:network} and~\ref{fig:tensor}).

The key idea in our method is to use this back projection tensor, $\mathbf{Z}$, as the input tensor to a deep convolutional neural network.  This approach converts a single sinogram into a stack of images each containing a collection of parallel lines of various intensities. Since each line is illustrated as constant intensity along its length, the information is stored in a highly redundant manner. However, the advantage is that the information is stored in a way that is spatially invariant, which is an important requirement in order for convolutional filters to be effective in a deep neural network. 

\subsection{Deep Learning for Image Reconstruction}
The goal of reconstruction is to find the mapping $F$ from the input sinogram $y$ to the latent clean image $x$ (e.g., $x=F(y)$). In order to find this mapping, we use a deep convolutional neural network~\cite{he2016} to predict the reconstructed image from the back projection tensor $\mathbf{Z}$.

Suppose we have $K$ sinogram / clean image pairs denoted as $\{(y_k^{tr},{x}_k^{tr})\}_{k=1}^{K}$. For each sinogram, we generate a stack of single-view back projections and build a training database $\{(\mathbf{Z}_k^{tr},{x}_k^{tr})\}_{k=1}^K$. 
We then minimize the mean squared error between the desired clean images and estimated ones from sinogram.
\begin{equation}
l(\mathbf{\Theta})=\frac{1}{2K}\sum_{k=1}^K ||x_k^{tr} - F(\mathbf{Z}_k^{tr};\mathbf{\Theta})||^2_2,
\label{eq:loss}
\end{equation}
where $\mathbf{\Theta}$ represents the trainable weight parameters in a deep neural network.

The reconstruction mapping $F(\cdot ;\mathbf{\Theta})$ is defined using the layers of a convolutional neural network as shown in Fig.~\ref{fig:network}. 
First, we apply 64 convolutional filter kernels of size $3\times 3 \times m$, where $m=16$ for 16 view angles data, to generate feature maps.
We then apply rectified linear units (ReLU)~\cite{nair2010} for neuron activation. 
It is worth noting that we use 3-dimensional convolutional kernels to apply the convolution operator to the full stack of single-view back projections.
Second, we apply 64 filters of 3$\times$3$\times$64 convolution kernel for 15 layers. 
A batch normalization unit is added between each convolution and a ReLU to avoid an internal covariate shift during mini-batch optimization~\cite{ioe2015}. 
Finally, we apply 1 filter of with a 3$\times$3$\times$64 convolution kernel to generate an image from the feature maps. 

By feeding an unseen testing sinogram $y^{te}$ into the trained reconstruction mapping $F(\cdot ;\mathbf{\Theta})$, we can generate the reconstructed image $\hat{x}^{te}$ directly from $y^{te}$.

\section{Experimental Results}
\label{sec:results}

We validate our Deep Back Projection (DBP) algorithm using simulated data representing 16-view CT scans on  multi-grain structures. 
We generate 100 synthetic noise-free images with size of 64$\times$64. 
We then apply a Radon transform to the synthetic images to collect the sparse CT sinogram. 
We then form the back projection tensor for each sinogram. 
An example of this dataset is shown in Fig.~\ref{fig:tensor}. 
We dedicate 80 scans (subdivided into patches as described below) to train the deep neural network for image reconstruction and leave the other 20 scans for testing. 

\begin{figure}[!b]
\centering
\subfloat{\includegraphics[width=3.25in]{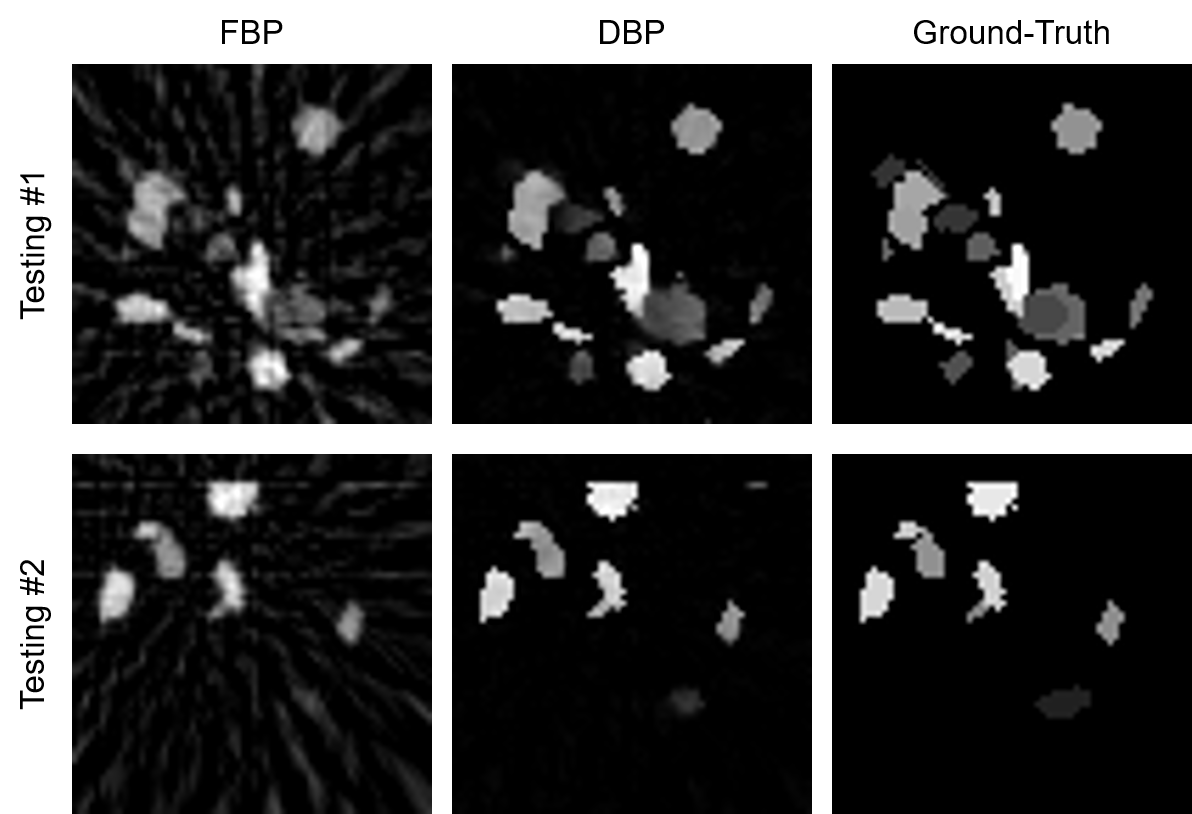}
}
\caption{Example of reconstruction results on 2 testing scans. Classical FBP shows spray artifacts due to incomplete CT data on limited view angles. Our DBP reduces the noise and enhances the spatial resolution compared with classical FBP, matching the image quality of noise-free ground-truth.}
\label{fig:result}
\end{figure}

A deep neural network is trained to learn the relation between single-view back projections and the clean ground-truth image from the same scan. 
To create the training database, 256000 patches of size of 8$\times$8 are extracted from 80 stacked backprojection/clean image pairs using data augmentation (e.g., horizontal and vertical flips, 90 degree rotations). 
It is worth noting that we extract patches from stacked back projections at the same location for all 16 view angles. We use the ADAM optimization~\cite{kingma2015} with gradually reduced learning rate from $10^{-3}$ to $10^{-5}$ with a total of 50 epochs. 
The size of a mini-batch is set to 128. 
The training procedures were implemented using the MatConvNet toolbox and took approximately 1 hour on a GTX TITAN X GPU.

Fig.~\ref{fig:result} shows an example of the image reconstruction results on the unseen testing scans. For comparison, we first reconstruct the classical filtered back projection (FBP) images illustrated in the left column of Fig.~\ref{fig:result}. 
We observe that classical FBP is susceptible to spray artifacts in the entire image domain when the CT scan data is acquired with limited view angles. 
In contrast, our DBP algorithm greatly reduces the noise and improves spatial resolution as depicted in the middle column of Fig.~\ref{fig:result}. 
For reference, we also display the ground-truth image in the right column of Fig.~\ref{fig:result}. 
In terms of texture, particularly inside the grain structure, the DBP reconstruction is very close to the noise-free ground-truth image. 
It is worth noting that the computational time for our DBP is under 10ms per slice, which is similar to that of classical FBP.  


\begin{table}[!t]
\caption{Quantitative Comparison between Reconstruction Results and Clean Ground-Truth Images}
\label{table:accuracy}
\begin{center}
\begin{tabular}{|c||c|c|c|}
\hline
&PSNR (dB) &SSIM\\
\hline
FBP & $18.43 \pm 3.75$ & $0.49 \pm 0.11$ \\
\hline
DBP & $19.84 \pm 2.44$ & $0.73 \pm 0.08$ \\
\hline
\end{tabular}
\end{center}
\end{table}

As a quantitative comparison, we report the peak signal to noise ratio (PSNR) and the structure similarity (SSIM) for 20 testing scans between reconstruction results and the clean ground-truth images in Table \ref{table:accuracy}. 
Our DBP outperforms classical FBP with about 1.4 dB PSNR value increase reflecting significantly reduced noise. 
In addition, our DBP significantly increases the SSIM value from 0.49$\pm$0.11 to 0.73$\pm$0.08 compared with the classical FBP. 
This indicates that multi-grain structures are better reconstructed in our DBP with high spatial resolution than classical FBP.  

\section{Conclusion}
\label{sec:conclusion}
In this paper, we present a deep learning method for sparse-view CT reconstruction directly from the sinogram data, which we call deep back projection (DBP).
Our method is based on the novel idea of constructing a ``back-projection tensor'' formed by a stack of single-view back projections.
Importantly, the back-projection tensor contains all the information in the original sinogram, but it results in a structure with spatial invariance so that the convolutional filters in a deep convolutional neural network (CNN) can be most effective.
We train a mutli-layer CNN to find the relationship between the back projection tensor and the clean ground-truth images, and we then use this CNN to perform tomographic reconstruction.
Results on a simulated 16-view CT scan show that our DBP method is very effective in reducing noise and enhancing the spatial resolution as compared to FBP. 
In addition, our DBP algorithm can preserve fine structures in the reconstructed image even with limited view angles, thus allowing faster acquisition without loss of quality.





\bibliography{DongHyeYe_Ref}{}
\bibliographystyle{IEEEbib}

\end{document}